\documentclass[conference]{IEEEtran}
\IEEEoverridecommandlockouts
\usepackage{cite}
\usepackage{amsmath,amssymb,amsfonts}
\usepackage{algorithmic}
\usepackage{graphicx}
\usepackage{textcomp}
\usepackage{xcolor}
\def\BibTeX{{\rm B\kern-.05em{\sc i\kern-.025em b}\kern-.08em
    T\kern-.1667em\lower.7ex\hbox{E}\kern-.125emX}}

\usepackage[utf8]{inputenc}		
\usepackage{listings}
 
\usepackage{tabularx}
\usepackage{url}
    
\usepackage{color}
    
\begin{document}

\title{Interoperability in the IoT - An Evaluation of the Semantic-Based Approach}

\author{\IEEEauthorblockN{Kristina Sahlmann}
\IEEEauthorblockA{Institute of Computer Science \\
\textit{University of Potsdam}\\
Potsdam, Germany \\
sahlmann@uni-potsdam.de}
\and
\IEEEauthorblockN{Florian Mikolajczak}
\IEEEauthorblockA{Institute of Computer Science \\
\textit{University of Potsdam}\\
Potsdam, Germany \\
florian.mikolajczak@uni-potsdam.de}
\and
\IEEEauthorblockN{Bettina Schnor}
\IEEEauthorblockA{Institute of Computer Science \\
\textit{University of Potsdam}\\
Potsdam, Germany \\
schnor@cs.uni-potsdam.de}
}

\maketitle

\begin{abstract}
While the management of heterogeneous network devices is usually solved by
protocols like SNMP and NETCONF, there is still no such accepted
solution for the management of heterogeneous IoT devices.
To avoid the vendor lock-in, several organizations like the IETF, W3C and ETSI
are working on standards with a semantic-based approach.
While the semantic approach seems to be appealing to solve
the interoperability problem, there is still the question whether
this approach is suited for constrained IoT devices.
Herein, we present the evaluation
of the MYNO, a semantic-based framework. MYNO is based on standards
and open-source libraries and aims to support the management
of constrained devices in the Internet of
Things. We demonstrate the benefits of the semantic-based
approach using a precision agriculture use case.
\end{abstract}


\section{Introduction}
\label{sec:Introduction}

Many vendor platforms for the Internet of Things (IoT) claim to be
interoperable. But such platforms~\cite{AWS.IoT, Azure.IoT}
provide vertical solutions for IoT systems: All components like IoT
devices, gateways, cloud-based services are from a single source. The
disadvantage is the dependency on the vendor for enhancements or
upgrades, the so-called \emph{vendor lock-in}. Connecting two vendor
solutions is near impossible.

Therefore, horizontal solutions are preferable. The idea is to
decouple the IoT components  from each other and use
common standards for communication
and data exchange. There are some initiatives for network management in
the IoT but still there are many open issues (i.e. interoperability,
scalability, security, energy saving) and no common comprehensive
solution~\cite{Silva.2019, Colakovic.2018}.

Moreover, most vendor platforms are cloud-based. If the connection to
the cloud is interrupted, local IoT devices might not operate
properly. This is not acceptable for Industrial IoT solutions or
solutions in the field of Smart Farming.  The solution is \emph{Edge
  Computing} where the managing edge node
is deployed in the local enterprise network
and 
accessible by devices even if the Internet connection is not
available. Furthermore, the latency is decreased and the network bandwidth
might be increased. If the edge node is connected to a cloud,
  for example for storing the data,
network bandwidth can be saved, when the amount of data is reduced
through pre-processing at the edge before transferring it to the
cloud.

A network in the IoT contains often hundreds of constrained devices
with sensors and actuators. Network configuration management tools are
required to maintain such a network.
One of the first IoT devices was a toaster in 1990,
connected to the Internet by TCP/IP and managed by {SNMP}~\cite{Romkey.2017}.
It had one control
operation to turn the power on and off. The authors wanted to
demonstrate the possibilities of SNMP. However, SNMP and the more recent
NETCONF
protocol~\cite{RFC.6241.Netconf} are not appropriated for
running on constrained devices~\cite{NETCONF-light.2012,Sehgal.2012}.

\begin{figure}[h]
\centering
\includegraphics[width=0.5\textwidth]{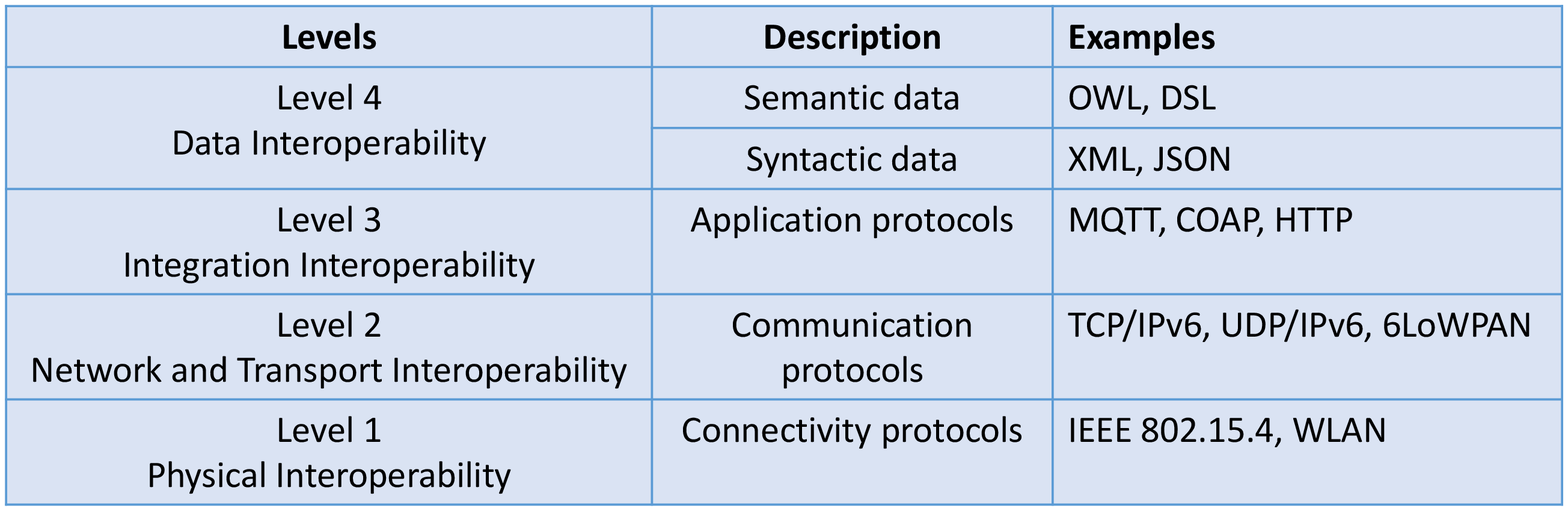}
\caption{Interoperability Model for the IoT}
\label{fig:interop-levels}
\end{figure}

Interoperability can be achieved by
using standards. However, {\em semantic} interoperability is necessary to
achieve full interoperability.
We propose an interoperability model, based on~\cite{Wang.2009} and adapted
for
the IoT, as shown in Figure~\ref{fig:interop-levels}. There can be
identified
four levels which need interoperability: (1) the physical level which
refers to the connectivity protocols like IEEE 802.15.4 or WLAN; (2)
the network and transport level addresses communication protocols like
TPC/IPv6, UDP/IPv6, 6LoWPAN; (3) the integration interoperability
includes application protocols like MQTT~\cite{MQTT.2019}, COAP~\cite{COAP.2014}, HTTP.
Obviously,  communication and application protocols are not
sufficient to achieve interoperability in an IoT
system. The meaning and the context of the exchanged data are also
necessary to be defined. This needs annotations like "what kind of device?": a
sensor, "which functionality?": provides measuring functionality, "which
kind of data?": for temperature data, "in which units?": in degree
Celcius or Fahrenheit. Such 
semantic annotations as wells as technical and non-technical
attributes for data can be provided by machine-interpretable descriptions~\cite{Barnaghi.2012}. This is addressed by level
(4) data interoperability which is subdivided into Syntactic Data
and Semantic Data. Syntactic Data describes the format and structure
 like XML and JSON; and Semantic Data describes the meaning of data
and the common understanding of vocabulary, e.g. with the help of
dictionaries, taxonomies, ontologies and formalization method for
sharing meaning with a model language, e.g. OWL~\cite{OWL.2012} or DSL~\cite{Vorto.2020}.

To tackle the interoperability problem, the MYNO framework
was proposed which is based on standards and open-source
implementations and acts on the edge of the
network~\cite{Sahlmann.Spain.2018}.  The MQTT protocol is the common
IoT protocol and the basis for the framework. The idea of the
solution is to bridge the network configuration protocol NETCONF~\cite{RFC.6241.Netconf} and the MQTT protocol and enhance the architecture
with semantics.
The proposed MYNO framework is based on the
following technologies: MQTT, YANG, NETCONF and Ontology, which build
the acronym. YANG~\cite{RFC.7950} is the data modeling language for the NETCONF
protocol. An ontology is used for semantic device-descriptions. The
MYNO framework pursues a semantic-based approach, supports
interoperability for heterogeneous IoT devices, and provides a
model-driven web client.

While the semantic approach seems to be appealing to solve the
interoperability problem, there is still the question whether this
approach is suited for constrained devices and thin edge
nodes like a Raspberry Pi.  Therefore, the contribution of this work
is the evaluation of the semantic-based approach for the IoT to answer
the following research questions:
\begin{itemize}
\item how are heterogeneous devices  supported by the MYNO framework? 
  (see Section~\ref{sec:MYNO} and ~\ref{sec:Use-Case}) 
\item how to implement semantic device descriptions on constrained
  devices? (see Section~\ref{sec:Use-Case})
\item which ontology should be used for device descriptions? (see Section~\ref{sec:MYNO})
\item how much resources are needed on the edge node? (see Section~\ref{sec:Evaluation})
\item does the system scale with the number of IoT devices? (see Section~\ref{sec:Evaluation})
\item is this effort justifiable? Advantages vs. disadvantages should be considered. (see Section~\ref{sec:Discussion})
\end{itemize}

The paper is structured as follows: First, we outline
the work at  standardization organizations for interoperability in the IoT.
Then, we present the semantic-based approach of the MYNO framework in Section~\ref{sec:MYNO}. The use case Precision Agriculture demonstrates in Section~\ref{sec:Use-Case} how new devices and sensors are integrated
in the framework. The performance evaluation is presented in Section~\ref{sec:Evaluation}. Finally, we discuss the results in Section~\ref{sec:Discussion}.

\section{Related Work}
\label{sec:Related-Work}

Organizations like the  IETF, W3C, ETSI and Open Mobile Alliance (OMA)
work on the standardization
of  an interoperability approach for the IoT~\cite{COAP.2014,WoT-Archi.2020,SAREF.2019,LWM2M.2019}.
Most of them use a semantic-based approach. However, they often do not present a holistic framework. For example, the IETF focuses on protocols like COAP~\cite{COAP.2014}, the CoAP Management Interface
(CORECONF)~\cite{CORECONF.2020}, and OMA on LwM2M~\cite{LWM2M.2019}. ETSI is a member of the oneM2M~\cite{oneM2M.FuncArchi.2019} initiative and developed the SAREF ontology~\cite{SAREF.2019} for the IoT. oneM2M provides a technical specification which claims to achieve interoperability solutions for M2M and IoT technologies. However, it is focusing on the middleware services and does not consider the underlying network and devices.

Recently, W3C finalized two recommendations for the Web of Things
(WoT): the WoT Architecture~\cite{WoT-Archi.2020} and the WoT Thing
Description (TD)~\cite{WOT-TD.2020}. The WoT is intended to enable
interoperability across IoT platforms and application domains. Similar
to the device descriptions in MYNO, the Thing Description is a
vocabulary which describes an IoT device and provides common concepts
for sensor observations, actions and events. They claim to follow the
syntax of JSON-LD, which is supposed to enable extensions and rich
semantic processing. However, the TD is a collection of many
vocabularies and not a valid ontology according to a Semantic Web
Standard like OWL. Therefore, semantic processing is not possible
yet. Further, the vocabulary of the TD does not provide descriptions of the communication protocols. Instead, Binding Templates~\cite{WTB.2020} were introduced to support different protocols like MQTT, CoAP, HTTP. However, from our experience, the vocabulary is not complete yet because some detailed constructs between protocols and device functionalities are missing. 

The WoT Architecture outlines many use case patterns.  A proprietary implementation is on the way\footnote{\url{https://projects.eclipse.org/projects/iot.thingweb}}. While WoT
 defines security and privacy aspects by design, the
 architecture stays very high-level and is missing
 important aspects for implementation, e.g. a discovery process. The
 MYNO framework
introduces a boostrapping process for discovery and has a
proof-of-concept implementation (see Section~\ref{sec:MYNO}).

The IETF also identified the advantages of semantics and started to work on a Semantic Definition Format (SDF)~\cite{Koster.2022} for data and interaction models in the IoT. The vocabulary is similar to the W3C TD and even smaller. It is using the JSON syntax and is not related to Semantic Web Standards. Therefore, the SDF has the same weaknesses: it is incomplete yet and semantic processing is not available. 

A recent IETF draft specifies a mapping between YANG, the
data modeling language for managed devices used by the network management protocol NETCONF, 
and the SDF format~\cite{Kiesewalter.2022}.
This is another indication that the MYNO approach is filling the gap, since it brings both worlds, network management and semantic, together.
The NETCONF-MQTT bridge in the MYNO framework does exactly this: it translate the semantic-based device descriptions to the YANG model for the NETCONF protocol. 

Obviously, there is a demand for semantic device descriptions and some  organizations try to specify an architecture using such descriptions. However, MYNO specified a holistic framework and provides an implementation.

\begin{figure*}[h]
\centering
\includegraphics[width=0.8\textwidth]{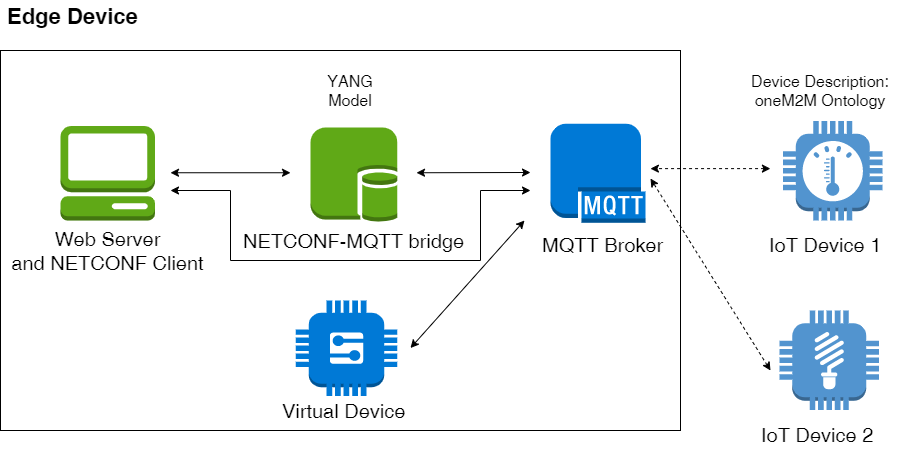}
\caption{MYNO System Architecture for Precision Agriculture}
\label{fig:myno-architecture-project}
\end{figure*}

\section{MYNO: A Semantic-based Approach}
\label{sec:MYNO}

This section gives a short overview over MYNO's components
and functionality. More details can be found in~\cite{Sahlmann.2021}.

The MYNO framework consists of four components shown in Figure~\ref{fig:myno-architecture-project}: a MQTT Broker, a NETCONF-MQTT bridge, a Web
Client, and a Virtual device.

The MQTT Broker is the central part for the MQTT communication which
is based on the publish-subscribe principle. Even constrained devices
like the Texas Instruments CC2538dk~\cite{TexasInstruments.2013}
board (see Table~\ref{tab:classes:examples}) 
with limited resources of computational power, memory, network
bandwidth, and energy, support this simple protocol. They can subscribe to
the so-called MQTT Topics and publish messages on such topics.

In the original network management architecture, a NETCONF server runs
  on the managed device and  responds to RPC calls. Since such a NETCONF server has shown to be too heavyweight for IoT devices~\cite{Sehgal.2012}, we proposed
the NETCONF-MQTT bridge which  provides a connection between the
NETCONF protocol and the IoT protocol, namely the MQTT protocol.

We introduced semantic device descriptions which provide the device
capabilities~\cite{Sahlmann.2017}.  The device descriptions are based
on the oneM2M Base Ontology~\cite{oneM2M.Ontology.2018}. This ontology was originally chosen among
many IoT ontologies for two reasons: (i) it is a small ontology for service and functionality description of devices which meets our requirements; and (ii) it is
represented by the OWL 2, a Semantic Web Standard.  Additionally, the SAREF ontology
from ETSI is related to the oneM2M ontology and developed vertical
domain ontologies. This might be a sign for a potential establishment
of this ontology. We extended the vocabulary of the ontology to
support the automatic generation of RPC calls by only four additional
OWL classes (\texttt{YangDescription, AutomationFunctionality,
  ConfigurationFunctionality, EventFunctionality}) and two OWL
Datatype Properties \texttt{mqttMethod} and \texttt{mqttTopic}. Using
an ontology model ensures that common device capabilities are reusable, machine-interpretable and can be used for rich semantic processing. 

A bootstrapping process defines how devices can join and leave a network managed by MYNO. This process has four CRUD operations and appropriate MQTT Topics. First, the devices publish their descriptions to the Topic \texttt{yang/config/create} during the create phase. The NETCONF-MQTT bridge
processes them and generates a YANG model with RPC calls for actuators
and descriptions for sensors. The YANG data model is used by the NETCONF protocol. On behalf of the NETCONF
protocol, the IoT network at the edge, can be managed as a part of the
entire enterprise network.

\begin{figure*}[h]
\centering
\frame{\includegraphics[width=0.8\textwidth]{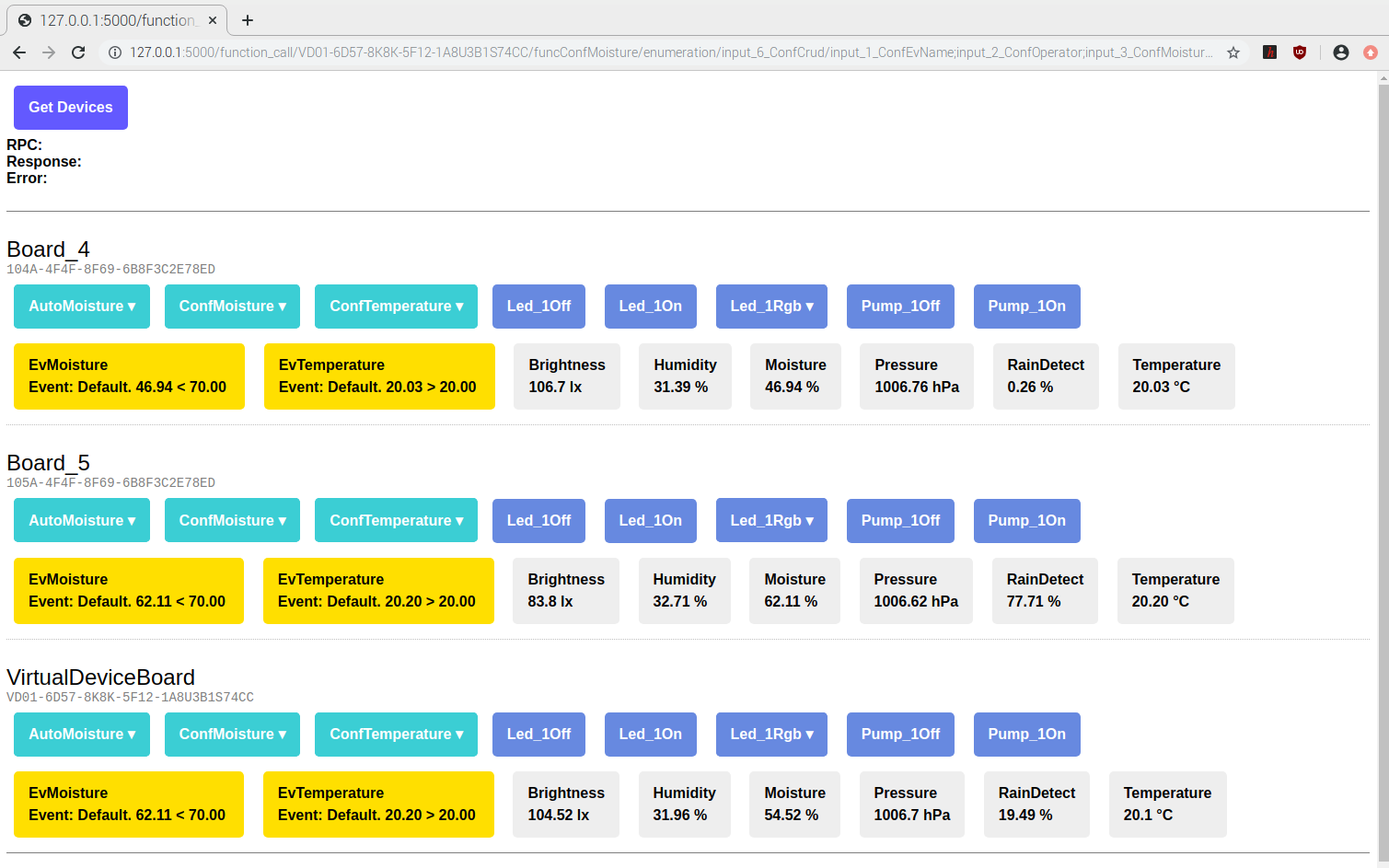}}
\caption{Web-Client for Precision Agriculture}
\label{fig:ptdw-web-client}
\end{figure*}

The Web Server acts as a NETCONF client and provides an graphical
interface based on the generated YANG model (see
Figure~\ref{fig:ptdw-web-client}). This is a model-driven approach. A
user can see which IoT devices are on the network and which
capabilities they have.  The grey fields represent the sensor
values. The blue buttons are triggers for the actuators with
parameters. The cyan buttons provide configurations of thresholds
(if-then-conditions) which can trigger events, shown in yellow fields.

A virtual device is an optional component on the edge and
enables the aggregation of IoT devices and sensor messages. This simplifies the
implementation
of applications like {\em ``give me the mean temperature of all rooms on
the south side''}.
The virtual device is started on the edge node and subscribes to
all bootstrapping topics and analyzes the device
descriptions to collect controlling and measuring functions as well as
configuration and automation functions. The virtual device publishes
its own device description to the MQTT broker and appears as a managed
device in the bridge. This way, the virtual device integrates
seamlessly into the MYNO architecture.

The MYNO source code is available as an open-source project\footnote{\url{https://github.com/ksahlmann/myno}}.

\subsection{Heterogeneous Devices}

\begin{table}[h]
\centering
\caption{Classification of IoT Devices}
\label{tab:classes:examples}
\begin{tabular}{|l|p{6cm}|}
\hline
Name & Example Devices (RAM, ROM/Flash)\\
\hline
Class 0 & -- \\ 
\hline
Class 1 & CC2538dk (32 KB, 512 KB) \\
\hline
Class 2 & ESP-32 NodeMCU (520 KB, 16 MB) \\
\hline
Beyond Class 2 & Arduino Yún Rev 2 (16 MB, 64 MB),\\
               & Raspberry Pi 3B (1 GB, 16 GB), Raspberry Pi Zero w (512 MB, 16 GB) \\
\hline
\end{tabular}
\end{table}

The feasibility of our approach with proposed scenarios and
heterogeneous devices in terms of capabilities (sensors, actuators, etc.) and constraints was
shown by several implementations. For new kinds of sensors or actuators, only the ontology-based device description was extended at minimum required. 
Since the YANG model is automatically generated
from the device descriptions on the edge node, no additional apps
from vendors are required. On the edge node, only the MYNO components are running: the bridge, the MQTT broker and optionally the virtual device. This approach is similar to the single source of
truth (SSOT) architecture where data are only managed at one place. In
MYNO, adding a new IoT device or new sensors requires only the update
of one source, namely the update of the MYNO bridge source, but no source code
from $n$ different vendors has be installed on the edge node. This is different from smart home solutions where either we have a
vendor lock-in or have to install apps from different vendors to manage a device.

Currently, MYNO supports
implementations for the Texas Instruments
CC2538dk board~\cite{Sahlmann.2020}, the Ard\'uino
Yún~\cite{Nowak.2018} and the ESP-32 NodeMCU. Table~\ref{tab:classes:examples} classifies
the devices regarding RFC~7228~\cite{RFC.7228} based on their resources.
The low-cost ESP-32 NodeMCUs were chosen for the scalability study
(see Section~\ref{sec:Evaluation}). The integration of these boards
within the MYNO framework is discussed in the next section to
demonstrate the usability of the MYNO approach.

\section{Use Case Precision Agriculture}
\label{sec:Use-Case}

For the evaluation of the MYNO framework, we setup 
a prototype implementation for IoT-based Precision Agriculture in a
greenhouse. Precision Agriculture is made possible by the
IoT
and is an ongoing research field~\cite{Ma.2011, Bauer2017, Marcu.2019,Chowdhury2020}.

The requirements are:
sensing environment data (air and soil), controlling irrigation, event
configuration and notification when thresholds are reached, and
finally, the automation of controlling functions (if-then condition).

\subsection{Testbed}
The prototype contains an edge component, a Raspberry Pi 3B, and 10 microcontroller boards which monitor 10 plants on the edge network. A single plant is representing a greenhouse or a field. 

A WLAN hotspot is installed in the lab as an access point for the
Raspberry Pi and the devices.
As shown in Figure~\ref{fig:myno-architecture-project}, the Raspberry
Pi has running the MQTT broker from Mosquitto, the NETCONF-MQTT bridge
and the web-based NETCONF Client as well as a Virtual Device.

The microcontroller boards are based on the low-priced ESP32 NodeMCU
Module\footnote{\url{https://www.az-delivery.de/products/esp32-developmentboard}}.
Every EPS32 board was extended through a breadboard equipped with
sensors and actuators.

The following sensors are wired with the breadboard:

\begin{itemize}
\item A capacitive soil moisture sensor v1.2 determines the dielectric constant of the soil which is an indicator for dry or wet soil. 
\item Three sensors, namely temperature, humidity and air pressure, are combined in a GY-BME280 module which  measures air condition.
\item A raindrops sensor measures the conductivity of its surface.
This is transformed into a binary output (rain/no rain) using an adjustable threshold.
\item A GY-302 BH1750 light sensor measures intensity of visible light in lux.
\end{itemize}

The following actuators are deployed on the breadboard:
\begin{itemize}
\item A 5V mini water pump with external power supply (2 AA batteries) and watering pipe is controlled through the relais.
\item A 1-relais 5V KY-019 module controls the water pump.
\item A KY-016 RGB LED module is used for state signaling like a traffic light.
\end{itemize}

\begin{figure}[h]
\centerline{\includegraphics[width=0.5\textwidth]{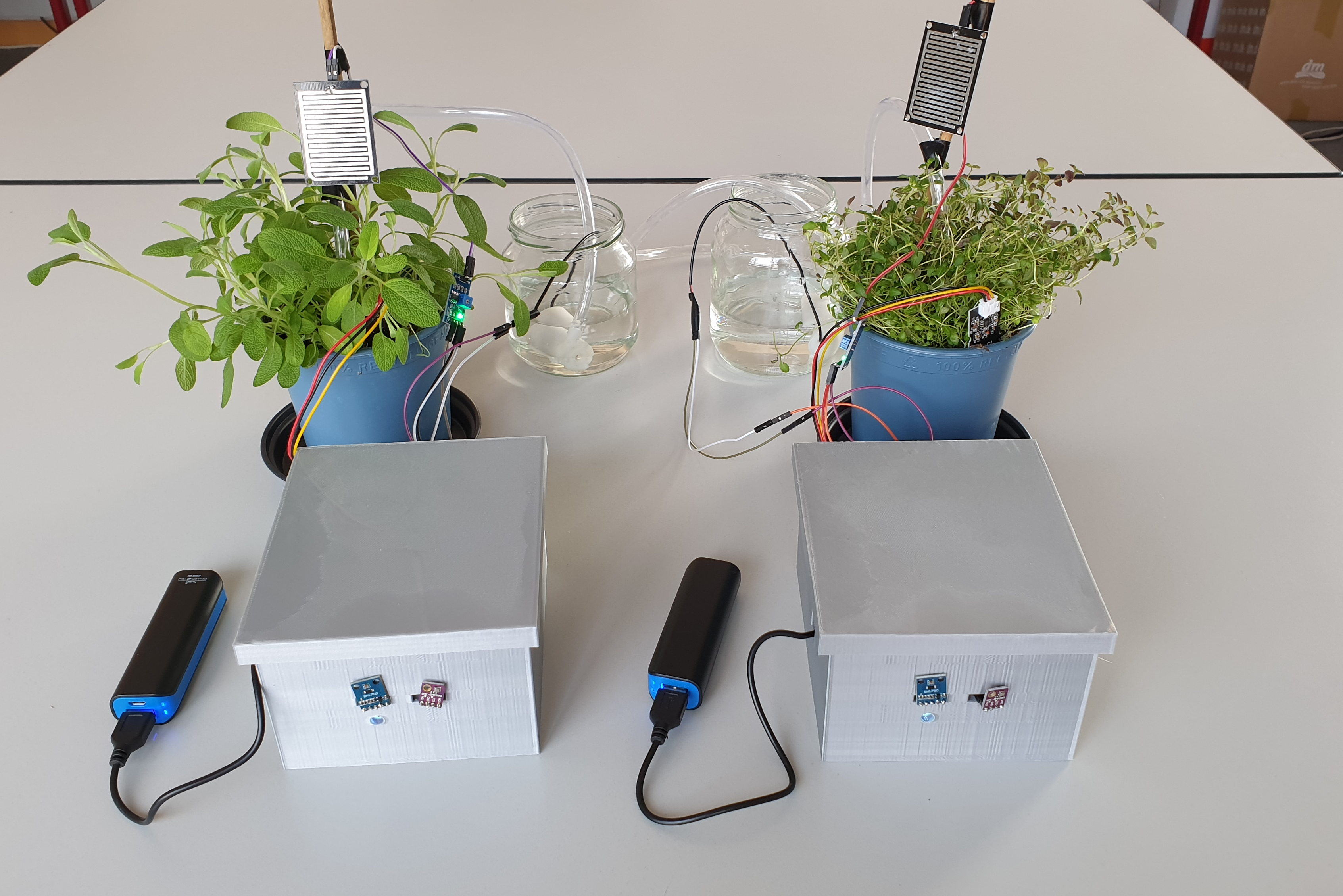}}
\caption{Two devices of the testbed. Well visible are also 
    the raindrop     sensors and the water pumps.}
\label{fig:agrarprojekt}
\end{figure}

The power supply for a EPS32 board is ensured through a powerbank
connected over the micro USB port. Figure~\ref{fig:agrarprojekt}
depicts two of the IoT devices in our testbed.

\subsection{Device Description}
  
\begin{figure}[h]
\centering
\includegraphics[width=0.5\textwidth]{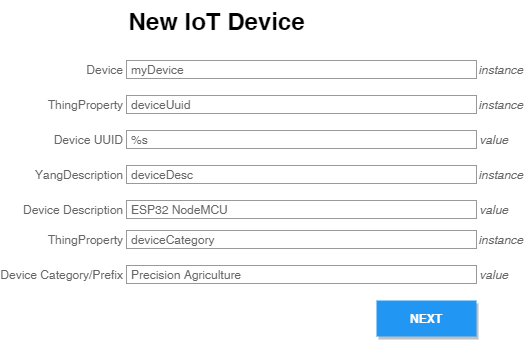}
\caption{Mockup Design for a new IoT Device}
\label{fig:device-editor}
\end{figure}

Usually, we use the Protégé~\cite{Protege.2016} tool for editing the semantic device descriptions. For simplicity, we designed some mockups for the web-based interface which could support manufacturers to create new device descriptions. Figure~\ref{fig:device-editor} shows a form for a new IoT device. There are some instances of OWL classes and values to define: a name for the Device; a Thing Property for the UUID (a unique ID for each device) and a value (or a placeholder); a name for the Yang Description and a value for Device Description as well as a name and a value for Device Category (all necessary for the YANG model). 

\begin{figure}[h]
\centering
\includegraphics[width=0.5\textwidth]{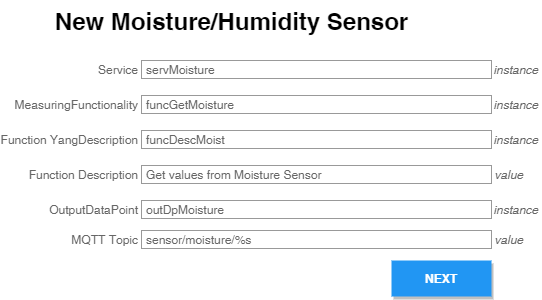}
\caption{Mockup Design for a new Sensor Function}
\label{fig:moisture-sensor-editor}
\end{figure}

Figure~\ref{fig:moisture-sensor-editor} shows a form for a new moisture sensor. This works straightforward according to the ontology model: a device has an instance of the Service and a Measuring Functionality. A Yang Description and a Function Description must be defined. An instance of the Output Datapoint tells that there is a data delivery endpoint. The MQTT Topic sets the actual Topic. 

A device description must contain all capabilities which are provided by such an agriculture device. The following controlling functions for actuators are defined:
\begin{enumerate}
\item switch the RGB LED on and off;
\item switch the RGB LED with a given RGB color;
\item turn the water pump on and off; 
\end{enumerate} 

For sensors, the device description was extended by reusing the OM-2 ontology~\cite{OM-2.2020} to provide units of measurements. The following sensor measurements are defined in the device description:
\begin{enumerate}
\item soil moisture in percent;
\item brightness in lux;
\item air humidity in percent;
\item air pressure in hectopascal;
\item air temperature in degree Celsius;
\item raindrops detection in percent; 
\end{enumerate}

For event configuration and notification, the device description was extended by by new OWL classes. Such configuration defines a threshold value for a sensor as well as an interval and duration for an event notification. Additionally, a name and a CRUD operation for this configuration must be defined. The difference to the controlling function is not only in the parameters which are always the same but also an MQTT Topic for publication of events like sensor values. The device description is  reusing the TIME ontology~\cite{W3C.Time.2020} to provide ontology classes for interval and duration. The configuration functions are defined for two critical sensor measurements: soil moisture and air temperature. For example, events should be published every 10 seconds during the next 60 seconds when the soil moisture is under 30 percent. 

The automation function (if-then condition) is a special case of event configuration and is defined in the device description as a combination of a configuration function and a controlling function instead of event notification. For example, if the soil is dry then turn the water pump on or switch the RGB LED to red to send a visual signal. Such automation functions can be used for event-based processing on a device instead of the event-based processing on the edge or in a cloud.

\section{Evaluation}
\label{sec:Evaluation}
Potentially, the MQTT broker is scalable but the NETCONF-MQTT bridge
behind must be fast enough to process device descriptions, sensor messages and RPC calls. We evaluated the scalability of the MYNO framework in our
testbed with up to 10 devices.

In our experiment, we follow the guidelines from
Jain~\cite{Jain.1991}. The scalability evaluation considers only messages with
device descriptions and sensor values. The actuator, configuration and
automation messages are not considered because their occurrence is
marginal.

The system parameters for the edge node,
a Raspberry Pi 3B, and the ESP32 NodeMCU Module are shown in Table~\ref{tab:classes:examples}.
On the Raspberry Pi 3B, Raspbian OS 3.4 was installed,
and the ESP32 NodeMCU devices were running with Arduino Sketch v. 1.8.10.
All system components communicate through WLAN on 2.4 GHz basis. 
Each ESP32 device has 6 sensors which publish periodically their data.
The sensors and their sensing interval are shown in
Table~\ref{tab:sensor-interval}.

\begin{table}[h]
\centering
\caption{ Sensing intervals}
\begin{tabularx}{0.46\textwidth}{|l|r|X|}
\hline
Sensor & Interval & MQTT Topic\\
       & [s]      & \\
\hline
Soil Moisture & 600 & \url{sensor/moisture/moisture_1/UUID} \\
\hline
Raindrops & 300 & \url{sensor/rain/rain_1/UUID} \\
\hline
Brightness & 60 & \url{sensor/brightness/brightness_1/UUID} \\
\hline
Air Temperature & 60 & \url{sensor/temperature/temperature_1/UUID} \\
\hline
Air Pressure & 60 & \url{sensor/pressure/pressure_1/UUID} \\
\hline
Air Humidity & 60 & \url{sensor/humidity/humidity_1/UUID} \\
\hline
\end{tabularx}
\label{tab:sensor-interval}
\end{table}

The evaluation is divided into three sub-experiments:
Device Description, Sensor Messages, and Energy
Experiment; and answers the following questions:

\begin{enumerate}
\item Is the edge node capable to process the device description
  within acceptable time?
\item Does the MYNO framework scale with an increasing number of device descriptions and sensor messages? 
\item Is the semantic approach feasible for constrained devices:
  What are the consequences in term of energy consumption?
\end{enumerate}

\subsection{Device Description Experiment}

The goal of this experiment is to investigate the scaling of the MYNO
framework during bootstrapping, when the  devices publish their device
descriptions. The processing of the semantic descriptions is implemented by
the RDFLib v. 4.2.2 library~\cite{RDFLib.2017} and this is the most
  computationally intensive part.

In this experiment, we vary the number of connected devices:
1, 3, 6, and 10 devices.  The experiment for each number of devices was
repeated three times. 

The measured  metrics are: 
\begin{itemize}
\item time for processing in the bridge,
\item CPU and RAM usage in the Raspberry Pi (\texttt{vmstat}),
\item time for transmission (\texttt{tshark}).
\end{itemize}

We turned on the devices at 60~s intervals one after another.
Since two devices are supplied from the same power bank, we started
the devices pairwise  in the case of 6 and 10~devices, again at 60~s
intervals.

\begin{table}[h]
\centering
\caption{Time for the RDFLib processing of a device description in the bridge in seconds (rounded)}
\begin{tabular}{|r|r|r|r|r|}
\hline
\# Dev. & AVG & MAX & MIN & MEDIAN\\
\hline
1  & 12.556165  & 12.778631  & 12.437580 & 12.452284 \\
\hline
3  & 8.551245 & 14.338671 & 5.430500 & 5.922269 \\
\hline
6  & 8.116861 & 14.419649 & 5.372922 & 7.516252 \\
\hline
10  & 7.564038  & 13.978273 & 5.624919  & 7.284434  \\
\hline
\end{tabular}
\label{tab:time-dd-processing}
\end{table}

Table~\ref{tab:time-dd-processing} shows the time for processing a
device description where the biggest part is the processing by the
RDFLib library.  The processing of the first device description takes
much longer than processing the following descriptions.  This behavior
is always observable after a restart of the bridge.  Obviously, there
is some initializing work done by the RDFLib before the first RDF querying.
But in the following, the median of the processing time  increases
only slightly from 5.9 up to 7.3~seconds.

\begin{figure}[h]
\centering
\includegraphics[width=0.5\textwidth]{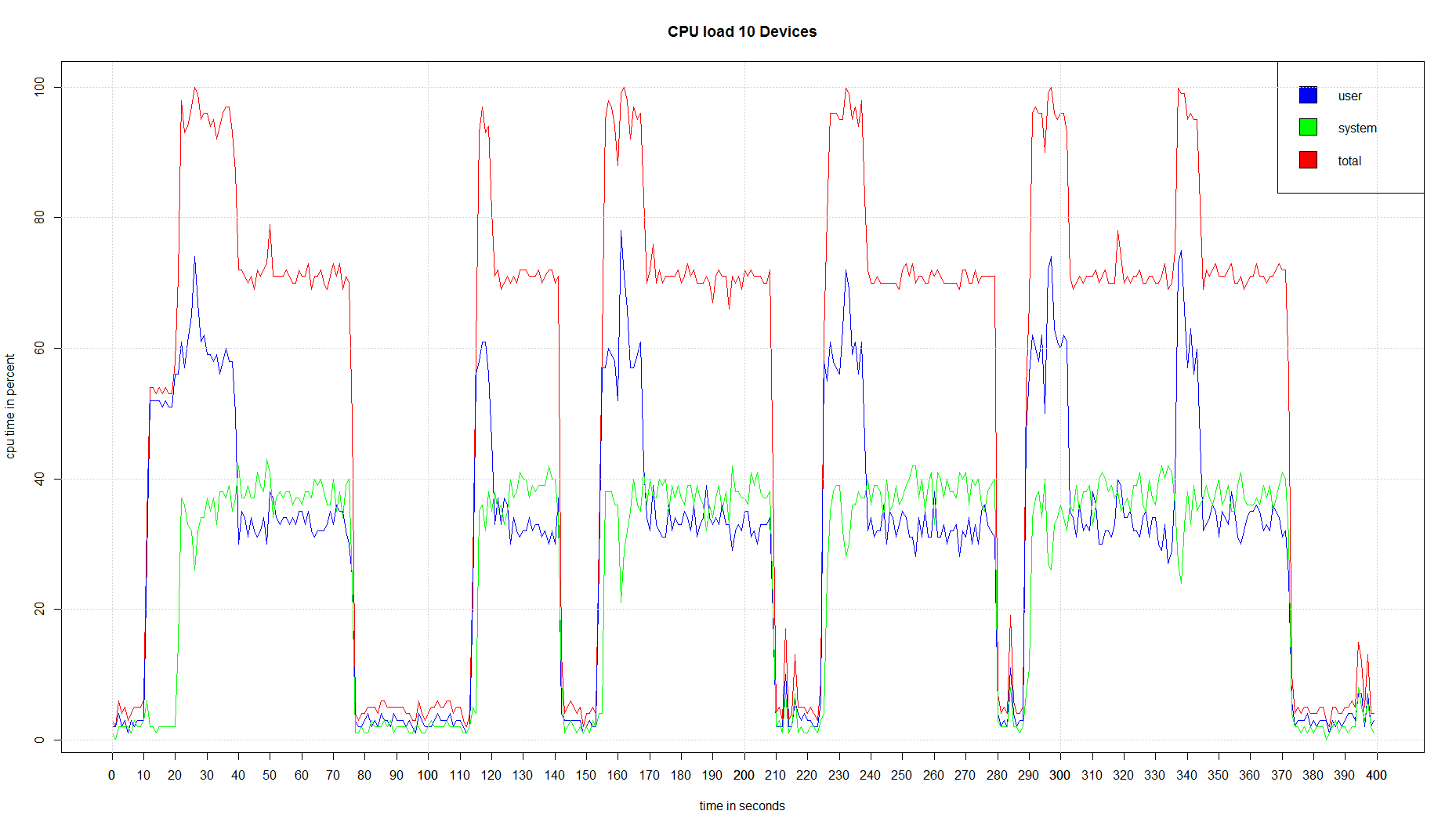}
\caption{CPU Load on the edge node: Starting 10 Devices.}
\label{fig:cpu-load-10device}
\end{figure}

The CPU load on the Raspberry Pi for starting 10 devices is shown in
Figure~\ref{fig:cpu-load-10device}. The CPU load increases drastically after
receiving a device description but after the  processing time it
decreases again to the previous level. The green line shows the time
spent running in kernel-mode, and the blue line shows the time spent
running in user-mode. Thus, obviously RDFLib performs some kernel tasks. There are also some short peaks which show that the processing of device
descriptions is a challenging task.

\begin{figure}[h]
\centering
\includegraphics[width=0.5\textwidth]{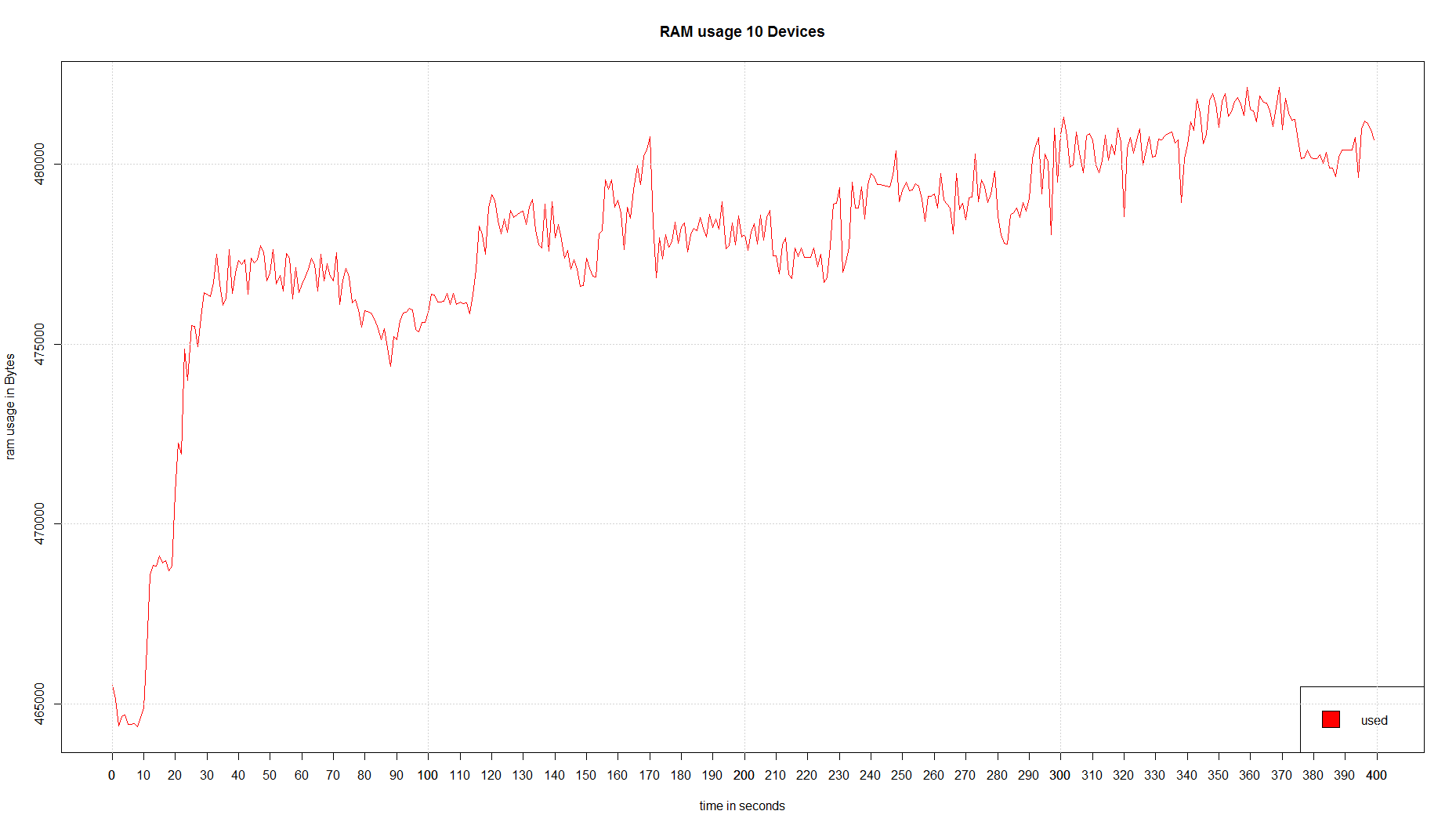}
\caption{RAM Usage on the edge node: Starting 10 Devices}
\label{fig:ram-usage-10d}
\end{figure}

Figure~\ref{fig:ram-usage-10d} shows the RAM usage which remains under
500~KB and increases
only slightly from 465 to 480~KB.

The transmission time of the device descriptions over
WLAN is fast, but the values vary strongly. We observed a minimum
transmission time of 26~ms (with 6 devices) and a maximum transmission time
of 809~ms (with 3 devices). Each running device sends every minute 4
messages with sensor data, every five minutes 5~messages and every ten
minutes 6~messages (see Table~\ref{tab:sensor-interval}).  So, the
influence of data collisions on the medium has a much higher influence
on the transmission time of the device descriptions
than the load on the edge node.

\subsection{Sensor Messages Experiment}

The goal of this experiment is to check whether the system scales with
  the number of sensor messages.  Therefore,  we measure
\begin{itemize}
\item the transmission  time of a sensor message, and
\item the number of retransmissions.
\end{itemize}

The interval between the sensor messages is shown in
Table~\ref{tab:sensor-interval}.  The sensor messages are MQTT
messages and consist of a topic and a sensor value. Hence, they are
comparable in their size. The runtime duration is 1 hour per experiment. 
Each experiment was repeated three times.

Even when 10~devices are running, we observed fast
transmission times between 74.4 
and 127.5 $\mu$seconds. Reasons for
these fluctuations are collisions on the medium
and retransmissions on TCP level.
Table~\ref{tab:sensor-messages} shows the number of retransmissions
monitored by the tshark tool running on the edge node.
The number of these so-called
{\em spurious}\footnote{https://www.chappell-university.com/post/spurious-retransmissions-a-concern} retransmissions is low and increases
almost linearly with  the increased number of devices.
Wireshark marks this kind of retransmissions as spurious  since
they are unnecessary: Wireshark has seen an ACK for the message, but
the sender retransmits it.
This is  most probably  a sign that the default timeout value within the
TCP communication stack on the ESP32 devices is too short in some rare cases.

\begin{table}[h]
\centering
\caption{Sensor Message Retransmissions}
\begin{tabular}{|r|r|r|r|}
\hline
\# Dev. & Messages Arrived &  Retrans.& [\%]\\
\hline
1  & 89  & 2  & 2.2\\
\hline
3  & 261 & 5  & 1.9 \\
\hline
6  & 521 & 11 & 2.1\\
\hline 
10 & 842 & 26 & 3.0\\
\hline
\end{tabular}
\label{tab:sensor-messages}
\end{table}

\subsection{Energy Experiment}

Energy efficiency is important for energy-constrained devices
i.e. powered by batteries.
The energy consumption in this experiment is measured by the lifetime of the fully charged
powerbanks (Schwaiger LPB220 533 powerbank with capacity of 2200 mAh).

Again, the sensors were configured as shown in
Table~\ref{tab:sensor-interval}.
We compare the lifetime of the devices for two different settings:
with and without sleeping state between the messages.
Without sleeping state,
the powerbanks lasted for 16~h 36~min. This means that the
boards shut down a part
of the hardware, wake up, measure, publish values and sleep again 30
seconds.  For this experiment, the lifetime of the powerbanks lasts
much longer, namely 1d 17h 51m.

Further, we optimized some processes in the MYNO framework concerning
the device descriptions to support
energy efficiency by design.  For example, the
size of the device description of the agriculture use case is 37~KB.
During bootstrapping and before sending
the device description, a device sends a request to the
NETCONF-MQTT bridge whether it is already registered on the
network. This is a much shorter message and
avoids an unnecessary transmission of the device description.

Additionally, the devices are configured with a compressed device
description to reduce the energy costs for sending the message.
In~\cite{Sahlmann.FGSN.2018}, the binary representations
RDF~HDT~\cite{HDT.2011} and CBOR~\cite{CBOR.2013} were evaluated for
the MYNO device descriptions. RDF HDT has shown much better space
savings than CBOR in our use case. We observed space savings of
72.68\% in RDF Turtle annotation and 84.06\% in RDF N-Triples
annotation because the input files have a verbose syntax. CBOR is not
well suited for the compression of ontologies, since long strings,
which are the main component of the device description, are not efficiently
compressed by CBOR (less than 15~\% savings in our example). But CBOR has of
course its strength when sensor data have to be encoded for
transmission.

     	   	
\section{Discussion of Results}
\label{sec:Discussion}

This section discusses the pros and cons of the MYNO approach.

\subsection{Interoperability}

Regarding the interoperability model for the IoT shown in
Figure~\ref{fig:interop-levels}, the results with the  MYNO prototype
are:

\begin{itemize}
\item Level 1: Physical Interoperability: demonstrated
  through implementations for IEEE 802.15.4~\cite{Sahlmann.2020} and WLAN.
\item Level 2: Network and Transport Interoperability: feasable through IP-based networks and protocols like IPv4/IPv6, 6LoWPAN~\cite{Sahlmann.2020} and TCP.
\item Level 3: Integration Interoperability: achieved through application protocols MQTT and NETCONF. 
\item Level 4: Data Interoperability: achieved through OWL standard and ontology-based device descriptions, and the YANG model for the NETCONF protocol. 
\end{itemize}

\subsection{The burden of the semantic approach}

While sensor data are typically only few Bytes in size, the size of
a device description might be much bigger.  For example, the size
of the device description of the agriculture use case is 37~KB.

The performance evaluation on the precision agriculture use case shows
that the edge node, namely the Raspberry Pi, is capable to process the
semantic device-descriptions within acceptable time. Further, the MYNO
framework scales with increasing number of devices and therefore its
descriptions and sensor messages.

The experiments have shown that the semantic approach is feasible for
constrained devices also in terms of energy consumption. The
bootstrapping process prevents unnecessary transmissions of device
descriptions. Thus, in best case it will be transmitted only once when
a device joins a network.

The data transmission over MQTT in a WLAN network worked well as the
performance evaluation has shown. However, it is challenging to transmit
several kilobytes of data in a 6LoWPAN network. Hence, bigger messages must be sent in \emph{slices} because of the constraint network bandwidth and memory
on the device~\cite{Sahlmann.2020}.

\section{Conclusion}

We evaluated the semantic-based framework MYNO in the context
of a high-precision agriculture use case. 
This work demonstrates that the semantic-based approach is suited for
constrained IoT devices and also for an edge computing architecture.

Data interoperability was achieved through semantic device
descriptions and the use of the de facto standard application
  protocol MQTT. The additional value of the ontology-based approach
is the underlying model which represents the meaning of the data and
which is self-descriptive and machine-readable. The YANG model for the
management of new devices is automatically generated from the device
descriptions. A further benefit of this approach is shown by the
concept of the Virtual Device which is useful for the aggregation of
device capabilities and sensor messages.

In the current version, MYNO uses RPCs for the implementation of the
actuator operations to interact with the NETCONF client. In RFC~8040~\cite{RFC.8040},
RESTCONF is specified which offers a REST-based interface to provide the actuator operations.
Like the RPC variant, the REST API is defined in the YANG model. Hence, MYNO framework can be easily extended with a REST interface.

Our experience with the MYNO framework shows that an IoT architecture
for interoperability is an interdisciplinary project which requires
knowledge in distinct fields like communication protocols, wireless
sensor technologies, and semantics.

\section*{Acknowledgment}
This research was funded by the Faculty of Mathematics and Natural Sciences of University of Potsdam in Germany. The responsibility for the content of this publication lies with the
authors.

\bibliographystyle{IEEEtran}
\bibliography{myno,rfc,w3c}

\end{document}